\documentclass[twocolumn,showpacs,preprintnumbers,amsmath,amssymb,times]{revtex4}

\usepackage{graphicx}% Include figure files
\usepackage{dcolumn}% Align table columns on decimal point
\usepackage{bm}% bold math

\begin{document}
\title{Phase transition and hysteresis in scale-free network traffic}

\author{Mao-Bin Hu$^1$}\email{humaobin@ustc.edu.cn}
\author{Wen-Xu Wang$^2$}
\author{Rui Jiang$^1$}
\author{Qing-Song Wu$^1$}\email{qswu@ustc.edu.cn}
\author{Yong-Hong Wu$^3$}

\affiliation{$^{1}$School of Engineering Science, University of
Science and Technology of China, Hefei 230026, P.R.C \\
$^{2}$Nonlinear Science Center and Department of
Modern Physics, University of Science and Technology of China,
Hefei 230026, P.R.C\\
$^3$Department of Mathematics and Statistics, Curtin University of Technology,
Perth WA6845, Australia}

\date{\today}

\begin{abstract}
We model information traffic on scale-free networks by 
introducing 
the node queue length $L$ proportional to the node degree 
and its delivering ability 
$C$ proportional to $L$. 
The simulation gives the overall capacity of the traffic system 
which is 
quantified by a phase transition from free flow to 
congestion. 
It is found that the maximal capacity of the system results from 
the case of the local routing coefficient $\phi$ slightly larger
than zero, and we provide an analysis for the optimal value of $\phi$.
In addition, we report for the first time the fundamental diagram 
of flow against density, in which hysteresis is found, and thus 
we can classify the traffic flow with four states: 
free flow, saturated flow, bistable and jammed.
\end{abstract}


\maketitle

%\section{Introduction}
Complex networks can describe many natural and social systems 
in which lots of entities or people are connected by physical links
or some abstract relations. 
Since the discovery of small-world phenomenon by Watts and 
Strogatz \cite{WS}, appeared in Nature in 1998, and scale-free 
property by Barab\'{a}si and Albert \cite{BA} one year later in
Science, complex networks have attracted growing 
interest among physics community
\cite{BA2,BA3,Newman,Newman2,Boccaletti}. 
As pointed out by Newman, the ultimate goal of studying complex 
networks is to understand how the network effects influence many 
kinds of dynamical processes taking place upon networks \cite{Newman}.
One of the dynamical processes, traffic of information or data 
packets is of great importance to be studied for the modern society. 
Nowadays we rely greatly on networks such as communication,
transportation, the Internet and power systems, and thus ensuring 
free traffic flow on these networks is of great significance  
and research interest.
In the pass several decades, a great number of works on the 
traffic 
dynamics have been carried out for regular and random 
networks.
Since the increasing importance of large communication networks 
with scale-free property such as the Internet \cite{PS}, 
the traffic flow on scale-free networks has drawn more and more 
attention \cite{Sole,Arenas,Tadic,Zhao,Mukherjee,Guimera,Guimera2,
Echen,Wang,Wang2,Yin,YanGang,deMenezes,deMenezes2,Germano}.

Researchers have proposed some models to mimic the traffic 
on complex networks by introducing the random generation 
and the routing of packets 
\cite{Sole,Arenas,Tadic,Zhao,Mukherjee,Guimera,Guimera2}.
Arenas et al. suggest a theoretical measure to investigate the 
phase transition by defining a quantity \cite{Arenas}, 
so that the state of traffic flow can be classified to the 
free flow state and the jammed state, where the free flow state 
corresponds to the number of created and delivered packet 
are 
balanced, and the jammed state corresponds to the packets 
accumulate on the network. 

Many recent studies have focused on two aspects to control 
the congestion and improve the efficiency of transportation: 
modifying underlying network structures or developing 
better route searching strategies in a large network 
\cite{Kleinberg}. 
Due to the high cost of changing the infrastructure, 
the latter is comparatively preferable.
In this light, Echenique et al., Wang et al. and Yin et al.
suggest traffic models based on the local information or the 
local integration of static and dynamic information 
\cite{Echen,Wang,Wang2,Yin}.
Yan et al. propose a efficient routing strategy based on the knowledge 
of the whole topology \cite{YanGang}. 
They find that the efficient path results in the redistributing 
traffic loads from central nodes to other noncentral nodes, and 
the network capability in handling traffic flow is improved 
more than 10 times by optimizing the efficient path.

However, previous studies usually assumed that the capacity of 
each node, i.e., the maximum queue length of each node for holding 
packets, is unlimited and the node handling capability, that is 
the number of data packets a node can forward to other nodes each 
time 
step, is either a constant or proportional to the 
degree 
of each node.
But, obviously, the capacity and delivering ability of a node 
are limited and variates from node to node in real systems, 
and in most cases, these restrictions could be very 
important in triggering congestion in the traffic system.

Since the analysis on the effects of the node capacity and 
delivering ability restrictions on traffic efficiency 
is still missing, 
we propose a new model for the traffic dynamics of such 
networks by taking into account the maximum queue length $L$
and handling capacity $C$ of each node. 
The phase transition from free flow to congestion is well 
captured and, for the first time, we introduce the fundamental 
diagram (flux against density) to characterize the overall 
capacity and efficiency of the networked system. 
Hysteresis in such network traffic is also produced.

%\section{The Traffic Model}
To generate the traffic network, our simulation starts  with 
the most general Barab\'{a}si-Albert 
scale-free network model 
which is in good accordance with real observation of 
communication networks \cite{BA2}. 
In this model, starting from $m_0$ fully connected nodes, one node
with $m$ links is added at each time step in such a way that
the probability $\Pi_i$ of being connected to the existing node
$i$ is proportional to the degree $k_i$ of the node, i.e.
$\Pi_i={k_i \over  \Sigma_j k_j}$, where $j$ runs over all existing
nodes. 

The capacity of each node is restricted by two parameters: 
(1) its maximum packet queue length $L$, which is proportional to 
its degree $k$ (a hub node ordinarily has more memory):
$L=\alpha \times k$; 
(2) the maximum number of packets it can handle per time step: 
$C=\beta \times L$. 
Here $0 < \beta \leq 1$ simply shows that the maximum number of 
handled packets is less than the maximum packet queue length $L$.
Motivated by the previous models \cite{Sole,Arenas,Tadic,Zhao,Echen,Wang,Wang2}, 
the system evolves in parallel according to the following rules:

1. Add Packets - Packets are added with a given rate $R$ (packets 
per time step) at randomly selected nodes and each packet is given 
a random destination.

2. Navigate Packets - Each node performs a local search among its 
neighbors. 
If a packet's destination is found in its nearest 
neighborhood, 
its direction will be directly set to the target. 
Otherwise, its direction will be set to a neighboring node $h$ 
with 
preferential probability:
$P_h={k^{\phi}_h \over \Sigma_i k^{\phi}_i}$.
Here the sum runs over the neighboring nodes, and $\phi$ is an 
adjustable routing parameter in that the packets are more likely to be 
forwarded to high degree nodes when $\phi > 0$.
It is assumed that the nodes are unaware of the entire network 
topology and only know the neighboring nodes' 
degree $k_i$.

3. Deliver Packets -- At each step, all nodes can deliver at 
most 
$C$ packets towards its destinations and FIFO 
(first-in-first-out) 
queuing discipline is applied at each node.
When the queue at a selected node is full, the node won't accept 
any more packets and the packet will stay at the site and wait 
for the next opportunity to be forwarded.
Once a packet arrives at its destination, it will be removed from 
the system. 
As in other models, we treat all nodes as both hosts and routers 
for generating and delivering packets.

\begin{figure}
\scalebox{0.7}[0.7]{\includegraphics{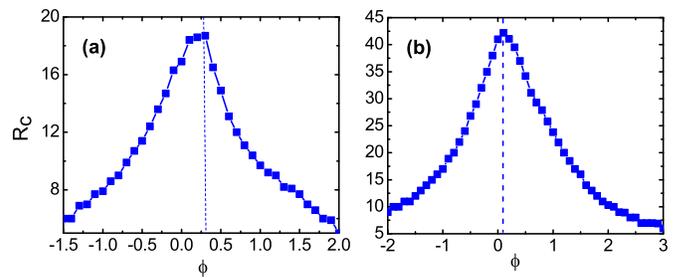}}
\caption{\label{Fig1}  (color online). The overall capacity of 
a network with $N=1000$, $m0=m=5$, $\alpha=1$(a), $\alpha=2$(b)
and $\beta=0.2$. 
The capacity is characterized by the critical value of $R_c$ for 
different $\phi$. 
In (a), $\alpha=1$, $\phi_{optimal}=0.3$ and $R_c^{max}=18.7$.
In (b), $\alpha=2$, $\phi_{optimal}=0.1$ and $R_c^{max}=42.2$.
In both cases, the maximum of $R_c$ corresponds to a $\phi$ slightly 
greater than zero marked by a dash line. 
The data are obtained by averaging $R_c$ over 10 network realizations.}
\end{figure}

We first simulate the traffic on a network of $N=1000$ nodes with 
$m0=m=5$. 
To characterize the system's overall capacity, we first 
investigate the increment rate $\eta$ of the number of packets 
in the system:
$\eta(R)=\lim_{t \rightarrow \infty} {\langle \Delta N_p \rangle 
\over \Delta t}$. 
Here $\Delta N_p = N_p(t+\Delta t)-N_p(t)$ with $\langle ... \rangle$ 
takes average over time windows of width $\Delta t$. 
Obviously, $\eta(R)=0$ corresponds to the cases of free flow state, which is 
attributed to the balance between the number of added and removed 
packets at the same time. 
As $R$ increases, there is a critical $R_c$ at which $N_p$ 
runs quickly towards the system's maximum packet number and $\eta(R)$
increases suddenly from zero, which indicates that packets 
accumulate in the system and congestion emerges and diffuses 
to everywhere. 

Hence, the system's overall capacity can be measured by the 
critical value of $R_c$ below which the system can
maintain its efficient functioning. Fig.\ref{Fig1}
depicts the variation of $R_c$ versus $\phi$. The maximum overall
capacity occurs at $\phi$ slightly greater than $0.0$ with
$R_c^{max}=18.7$ at $\phi=0.3$ for $\alpha=1$ (a) 
and $R_c^{max}=42.2$ at $\phi=0.1$ for $\alpha=2$ (b). 
The results are averaged from 10 simulations.

The analytical estimation of $R_c$ is too complicated for our 
routing model. 
In a recent paper \cite{Germano}, Germano and de Moura present 
analytical work on the rather simple traffic of particle hopping 
in complex networks.
In the following, we provide an analysis for the optimal value 
of $\phi$ corresponding to the peak value of $R_c$.
In the case of $\phi=0$, packets perform random-like walks if 
the maximum queue length restriction of each node is neglected.
The random walk process in graph theory has been extensively 
studied. 
A well-known result valid for our analysis is that the time the 
particle spends at a given node is proportional to the degree
of such node in the limit of long times \cite{Bollob}.
Similarly, in the process of packet delivery, the number of 
received packets (load) of a given node averaging over a 
period of time is proportional to the degree of that node. 
Note that the packets delivering ability $C$ 
of each node assumed to be proportional to its degree, 
so that the load and delivering ability of each node are 
balanced, which leads to a fact that no congestion occurs 
earlier on some nodes with particular degree than on others.
Since in our traffic model, an occurrence of congestion at 
any node will diffuse to the entire network ultimately, 
no more easily congested nodes brings the maximum network 
capacity. 
However, taking the maximum queue length restriction into 
account, short queue length of small degree nodes make them
more easily jammed, so that routing packets preferentially 
towards large degree nodes slightly, i.e., $\phi$ slightly 
larger than zero, can induce the maximum capacity of the 
system. 

This also explain the difference in the position of $R_c^{max}$
of Fig.\ref{Fig1}(a) and Fig.\ref{Fig1}(b). 
Comparing with the case of $\alpha=2$, the small degree nodes 
are more easy to jam when $\alpha=1$, so a greater $\phi$ is 
needed to achieve 
a more efficient functioning of the system. 
One can also conclude that the optimal $\phi$ will be zero if 
$\alpha$ is large enough. 

\begin{figure}
\scalebox{0.70}[0.70]{\includegraphics{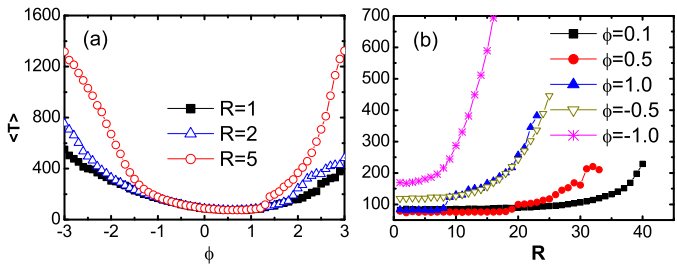}}
\caption{\label{Fig2}  (color online). Average travel time for 
a network with $N=1000$, $m_0=m=5$, $\alpha=2$ and $\beta=0.2$. 
(a) Average travel time $\langle T \rangle$ versus $\phi$ for 
$R=1$, $2$ and $5$. 
The data are truncated because the system jams when $\phi$ is 
either too large or too small. 
(b) The variation of $\langle T \rangle$ versus $R$ when $\phi$ 
is fixed. 
The data are also truncated when the system jams. }
\end{figure}

Then we simulate the packets' travel time which is also 
important for measuring the system's efficiency. 
In Fig.\ref{Fig2}(a), we show the average travel time 
$\langle T \rangle$ versus $\phi$ under the conditions of 
$R=1$, $2$ and $5$. 
In the free-flow state, almost no congestion on nodes occurs and 
the time for packets waiting in the queue is negligible, 
therefore, the packets' travel time is approximately equal to
their actual path length in map. 
But when the system approaches a jammed state, the travel time 
will increase rapidly. 
One can see that when $\phi$ is slightly greater than zero, the 
minimum travel time is obtained.
In Fig.\ref{Fig2}(b), the average travel time is much longer
when $\phi$ is negative than it is positive. 
These results are consistent with the above analysis that a 
maximum $R_c$ occurs when $\phi$ is slightly greater than zero. 
Or, in other words, this can also be explained as: 
when $\phi>0$, packets are more likely to move to the nodes with 
greater degree (hub 
nodes), which enables the hub nodes to be 
efficiently used and enhance the system's overall capability; 
but when $\phi$ is too large, the hub nodes will more probably 
get jammed, and the efficiency of the system will decrease.

%\section{Fundamental Diagram and Hysteresis}
Finally, we study the fundamental diagram of network 
traffic with our model. 
Fundamental diagram (flux-density relation) is one of the most 
important criteria that evaluates the transit capacity for a 
traffic system. 
Obviously, if the nodes are not controlled with the queue length 
$L$, the network system will not have a maximum number of packets 
it can hold and the packet density can not be calculated, so that 
the fundamental diagram can not be reproduced. 
%Interestingly, phase transition and hysteresis phenomena can be 
%observed in the fundamental diagram of our model.
%Our model reproduced the phase transition and hysteresis in 
%fundamental diagram. 
%Moreover, the fundamental diagram is different from a single-lane 
%traffic model or a BML-like model.

\begin{figure}
\scalebox{0.80}[0.80]{\includegraphics{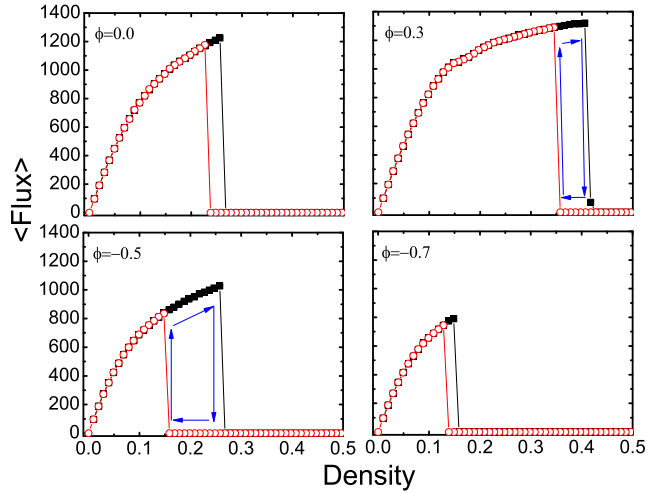}}
\caption{\label{Fig3} (color online). Fundamental diagram for a 
$N=1000$ network with $m0=m=5$, $\alpha=1$, $\beta=0.2$, 
and different $\phi$. 
The data are averaged over 10 typical simulations on one 
realization of network. 
In each chart, the solid square line shows the flux variation 
when adding packets to the system (increase density),
while the empty circle line shows the flux variation when drawing
out packet from the system (decrease density). 
The sudden transition density values are: 0.26 and 0.23 ($\phi=0.0$),
0.40 and 0.34($\phi=0.3$), 0.26 and 0.15($\phi=-0.5$), 
0.15 and 0.13($\phi=-0.7$).
For different realizations of network, the fundamental charts are 
similar, but with small difference in the transition values. 
The arrows in charts of $\phi=0.3$ and $-0.5$ are showing the 
hysteresis for guide of eyes.}
\end{figure}

To simulate a conservative system, we count the number of removed
packets at each time step and add the same number of packets to the
system at the next step. 
The flux is calculated as the number of successfully delivered 
packets from node to node through links per step. 
In Fig.\ref{Fig3}, the fundamental diagrams for $\phi=0.0,0.3,-0.5$
and $-0.7$ are shown. 

The curves of each diagram show four flow states: free flow, 
saturate flow, bistable and jammed. 
For simplicity, we focus on the $\phi=0.3$ chart with the maximum 
$\langle Flux \rangle=1319$ in the following description.
As we can see, when the density is low (less than $\approx 0.10$), 
all packets move freely and the flux increases linearly with 
packet density, which is attributed to a fact that in the free 
flow state, all nodes are operated below its maximum delivering 
ability $C$. 
Then the flux's increment slows down and the flux gradually 
comes to saturation ($0.10 \sim 0.34$), where the flux is 
restricted mainly by the delivering ability $C$ of nodes. 

At the region of medium density, the model reproduces an important 
character of traffic flow - ``hysteresis", which can be seen that 
two branches of the fundamental diagram coexist 
between $0.34$ and $0.40$. 
The upper branch is calculated by adding packets to the system, 
while the lower branch is calculated by removing packets from a 
jammed state and allowing the system to relax after the intervention.
In this way a hysteresis loop can be traced (arrows in Fig.\ref{Fig3}), 
indicating that the system is bistable in a
certain range of packet density. 
As we know so far, it is the first time that the hysteresis 
phenomenon is reported in the scale-free traffic system.

\begin{figure}
\scalebox{0.85}[0.85]{\includegraphics{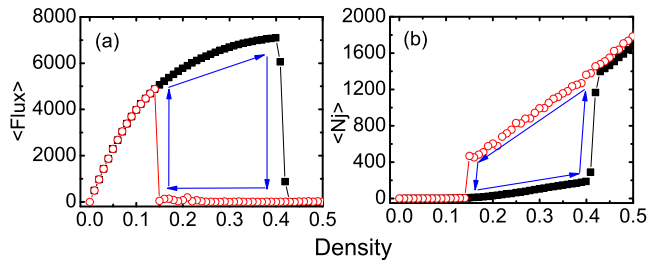}}
\caption{\label{Fig4} (color online). 
(a) Fundamental diagram for a $N=5000$ network with $m0=m=5$, 
$\alpha=1$, $\beta=0.2$ and $\phi=0.1$. 
(b) The averaged number of jammed nodes $\langle N_{jv}\rangle $. 
The symbols for increasing/decreasing density are the same as 
in Fig.\ref{Fig3}.
One can see that the two sudden change points $0.40$ and $0.14$ 
in both charts are equal.
The arrows are showing the hysteresis for guide of eyes.
}
\end{figure}

One can also notice that when $\phi=0.3$, the maximum saturated 
$\langle Flux \rangle$ is higher than others, and the saturated 
flow region is much boarder than the cases of 
$\phi=0.0,-0.5$ and $-0.7$.
All these results show that the system can operate better when 
$\phi$ is slightly greater than zero, which is also in agreement 
with the simulation result of $R_c$ in Fig.\ref{Fig1}.

In order to test the finite-size effect of our model, we simulate 
some systems with bigger size. 
The simulation shows similar phase transition and hysteresis in 
fundamental diagram as shown in Fig.\ref{Fig4}(a).

The flux's sudden drop to a jammed state from a saturated flow 
indicates a first order phase transition, which can be explained 
by the sudden increment of full (jammed) nodes in the system 
(See Fig.\ref{Fig4}(b)). 
According to the evolutionary rules, when a given node is full, 
packets in neighboring nodes can not get in the node. 
Thus, the packets may also accumulate on the neighboring nodes and 
get jammed. 
This mechanism can trigger an avalanche across the system when 
the packet density is high. 
As shown in Fig.\ref{Fig4}(b), the number of full nodes increase 
suddenly at the same density where the flux drop to zero and 
almost no packet can reach its destination. 
As for the lower branch in the bistable state, starting from 
an initial jammed configuration, the system will have some 
jammed nodes that are difficult to dissipate. 
Clearly, these nodes will decrease the system efficiency by 
affecting the surrounding nodes until all nodes are not jammed,
thus we get the lower branch of the loop.

%Moreover, an important conclusion can be drawn by comparing the
%charts in Fig.\ref{Fig3} that the $\phi=0.1$ chart has a much 
%broader bistable region than the $\phi=0.0$ one.
%This means, when the system retreats from a heavy load jammed state, 
%it is more difficult to reach a high efficiency state if $\phi$ is 
%greater than zero that packets are more likely to move to hub nodes. 
%In other words, though it is wise to take full advantage of the hub 
%nodes when the entire traffic is light, it won't be so happy to do 
%so at rush hours.

%\section{CONCLUSION}
In conclusion, a new model for scale-free network traffic is proposed 
to consider the nodes' capacity and delivering ability. 
In a systemic view of overall efficiency, the model reproduces 
several significant characteristics of network traffic, such as phase 
transition, and for the first time, the fundamental diagram 
for networked traffic system. 
Influenced by two factors of each node's capability and navigation 
efficiency of packets, the optimal routing parameter $\phi$ is found 
to be slightly greater than zero to maximize the whole system's 
capacity. 
A special phenomenon - the ``hysteresis" - is also reproduced in 
the typical fundamental diagram, indicating that the system is 
bistable in a certain range of packet density. 
It is the first time that the phenomenon is reported in networked 
traffic system.
For different packet density, the system can self-organize to four 
different phases: 
free-flow, saturated, bistable and jammed. 

Our study may be useful for evaluating the overall efficiency of 
networked traffic systems, and the results may also shed some light 
on alleviating the congestion of modern technological networks. 
%With different $\alpha$ and $\beta$, this model may also be applied 
%to other networks, such as power grid and even urban traffic systems 
%\cite{Rosvall,humaobin}. 

%\section*{ACKNOWLlinkMENT}
This work is funded by National Basic Research Program of China (No.2006CB705500),
the NNSFC under Key Project No.10532060, 10635040, Project Nos.70601026, 10672160, 
10404025, the CAS President Foundation, and by the Chinese Postdoctoral Research 
Foundation (No.20060390179).
Y.-H. Wu acknowledges the support of the Australian Research Council through 
a Discovery Project Grant.

\end{document}